\documentclass[aps,prd,preprint,preprintnumbers,unsortedaddress,superscriptaddress,showpacs,nofootinbib]{revtex4-1}

\usepackage{graphicx,color}
\usepackage{relsize}
\usepackage{slashed}
\usepackage{color}
\usepackage{ulem}
\usepackage{tabu}
\usepackage{bm}

\usepackage{amsmath}
\usepackage{amssymb}
\usepackage{amsthm}
\usepackage{mathrsfs}
\usepackage{graphicx}
\usepackage{epstopdf}
\usepackage{fancyhdr}
\usepackage{array}
\usepackage[all]{xy}
\usepackage{eufrak}
\usepackage{euscript}
\usepackage{enumerate}
\usepackage{slashed}
\usepackage{hyperref}
\usepackage{subfigure} 
\usepackage{epstopdf} 

\hypersetup{pdftex,colorlinks=true,linkcolor=blue,citecolor=blue,menucolor=black,urlcolor=blue,filecolor=blue}

\newcommand{\tr}{{\rm Trace}}
\newcommand{\minv}{M_{\rm inv}}

\newcommand{\br}{{\rm Br}}

\newcommand{\mev}{{\rm MeV}}
\newcommand{\tento}[1]{\times 10^{#1}}

\renewcommand\sout{\bgroup \color[rgb]{1,0,0} \ULdepth=-.5ex \ULset}

\begin{document}

\title{Theoretical description of the $\boldsymbol{J/\psi \to \eta (\eta') h_1(1380)}$, $\boldsymbol{J/\psi \to \eta (\eta') h_1(1170)}$ and $\boldsymbol{J/\psi \to \pi^0 b_1(1235)^0}$ reactions}

\author{Wei-Hong Liang}
\email{liangwh@gxnu.edu.cn}
\affiliation{Department of Physics, Guangxi Normal University, Guilin 541004, China}
\affiliation{Guangxi Key Laboratory of Nuclear Physics and Technology, Guangxi Normal University, Guilin 541004, China}

\author{S. Sakai}
\email{shsakai@itp.ac.cn}
\affiliation{Institute of Theoretical Physics, CAS, Zhong Guan Cun East Street 55 100190 Beijing, China}

\author{E.~Oset}
\email{oset@ific.uv.es}
\affiliation{Department of Physics, Guangxi Normal University, Guilin 541004, China}
\affiliation{Departamento de F\'{i}sica Te\'{o}rica and IFIC, Centro Mixto Universidad de Valencia - CSIC,
Institutos de Investigaci\'{o}n de Paterna, Aptdo. 22085, 46071 Valencia, Spain}

\begin{abstract}
 We have made a study of the $J/\psi \to \eta' h_1, \eta h_1$ (with $h_1$ being $h_1(1170)$ and $h_1(1380)$) and $J/\psi \to \pi^0 b_1(1235)^0$ assuming the axial vector mesons to be dynamically generated from the pseudoscalar-vector meson interaction. We have taken the needed input from previous studies of the $J/\psi \to \phi \pi \pi, \omega \pi \pi$ reactions.
 We obtain fair agreement with experimental data and provide an explanation on why the recent experiment on $J/\psi \to \eta' h_1(1380), h_1(1380) \to K^{*+} K^- +c.c.$ observed in the $K^+ K^- \pi^0$ mode observes the peak of the $h_1(1380)$ at a higher energy than its nominal mass.
\end{abstract}



\maketitle

\section{Introduction}
\label{sec:intro}

The axial vector mesons of low energy are emerging as a powerful source of information on hadron dynamics.
In quark models \cite{isgur,vijande}, they are not so well reproduced as the corresponding vector mesons.
Subsequent studies in the context of QCD, large $N_c$ behavior, combined with phenomenology,
have shown that the vector mesons are largely $q\bar q$ objects \cite{Pelaez}.
However, this is not the case following the same investigation, for the low-lying axial vector mesons \cite{rocaram}.
Actually these axial vector mesons have been studied within the context of the chiral unitary approach and have proved to be well described from the interaction of pseudoscalar and vector mesons \cite{Lutz,Luis,geng,Leupold} using potentials provided by the chiral Lagrangians \cite{Birse}.
Radiative decays of the axial vector mesons have been largely studied from different points of view \cite{rocahosa,hiderroca,yongseok},
and more concretely from the point of view as being dynamically generated from the pseudoscalar-vector interaction \cite{Lutzleo,hiddenaga}.
One of the topics where axial vector mesons are shown to play an important role is in $\tau^-$ decays.
The $\tau^- \to \nu_\tau \pi^+ \pi^- \pi^-$ decay has been shown to be dominated by the formation of the $a_1(1260)$ using different models \cite{Roig,Volkovtau,Osipov}, and concretely assuming the resonance to be largely made from the $\rho \pi$ interaction \cite{Leupold}.
Another $\tau^-$ decay more recently studied is the $\tau^- \to \nu_\tau \pi^- f_1(1285)$ which is investigated in Ref. \cite{Volkovtauf1} from the perspective of the Nambu--Jona-Lasinio model and in Ref. \cite{luistauf1} from the perspective of  the $f_1(1285)$ being dynamically generated from the $K^* \bar K$ interaction.
A generalization of this reaction to the general case of $\tau^- \to \nu_\tau P A$  ($P$ for pseudoscalar and $A$ for axial vector meson) has been done in Ref. \cite{Dairoca}
and estimates based on vector meson dominance have been done before \cite{calderon}.
Other weak decays, as the $B_s^0 \to J/\psi f_1(1285)$, have also been studied \cite{Molina}.

Strong decays of axial vector mesons have also been investigated. The three pion decay of the $a_1(1260)$ is studied in Ref. \cite{ZhangXie},
the strong decays of the $b_1(1235)$ in Ref. \cite{yongseok}, and the decay of $f_1(1285) \to \rho^0 \pi^+ \pi^-$ has been investigated in Ref. \cite{OsiVolkov}.
The $f_1(1285)$ is a particular case in chiral theory since it is generated from the single channel $K^* \bar K -c.c.$ in $s$-wave \cite{Luis,geng}.
The $K\bar K \pi$ decay of the $f_1(1285)$ is investigated in Ref. \cite{Aceti} and the experimental rates \cite{PDG} are well reproduced.
More concretely, the $f_1(1285) \to \pi^0 a_0(980) (\pi^0 f_0(980))$ decays have been addressed in Ref. \cite{Acetiso}.
These reactions pose the double challenge of treating both the $f_1$ and the $f_0, a_0$ resonances as dynamically generated,
the $f_0, a_0$ from the pseudoscalar-pseudoscalar interaction \cite{Oller,Kaiser,Markushin,juan}.
The $\pi^0 a_0(980)$ decay mode is well reproduced and, in addition, predictions are made for the isospin-forbidden $\pi^0 f_0(980)$ mode.
It is interesting to note that this decay mode was later confirmed by experiment in Ref. \cite{BESiso}, in agreement with the strength and the narrow shape predicted in Ref. \cite{Acetiso}.
Other strong or electromagnetic reactions involving axial vector mesons include the $\gamma p \to \pi^+ \pi^+ \pi^- \eta$ reaction studied in Ref. \cite{ZhangXie},
the $\pi^- p \to a_1(1260)  p$ reaction studied in Ref. \cite{cao},
the $e^+ e^- \to f_1(1285) \gamma (a_1(1260) \gamma)$ reactions studied in Ref. \cite{Volkovee}, and the $K^- p\to f_1(1285) \Lambda$ reaction studied in Ref. \cite{XieLam}.
The $f_1(1285)$ has also been observed in the CLAS experiment \cite{CLASf1} and studied theoretically in Ref. \cite{He}.

$J/\psi$ decays have also contributed to this field and studies have been done for the $J/\psi \to \phi K\bar K^*$ and $J/\psi \to \phi f_1(1285)$ decays \cite{Jujun}.
Another interesting reaction is the $J/\psi \to \eta K^{*0} \bar K^{*0}$ reaction measured by the BES collaboration in Ref. \cite{besalba},
where the peak observed in the $K^{*0} \bar K^{*0}$ invariant mass around the $K^{*0} \bar K^{*0}$ threshold was interpreted in Ref. \cite{albala}
as a manifestation of the $h_1(1830)$ axial vector meson predicted by Ref. \cite{GengOset}.
The behavior of $a_1(1260)$ resonance in a nuclear medium from the perspective of its $\rho \pi$ main component was also studied in Ref. \cite{Cabrera}.

Very recently the BESIII collaboration has reported the observation of the $h_1(1380)$ in the $J/\psi \to \eta' K\bar K \pi$ decay \cite{Besnew}.
Selecting $\eta$ or $\eta'$ in the final state is a good filter for isospin and $C$-parity such that the extra meson produced is a $I^G(J^{PC})=0^- (1^{+-})$ state,
assuming $s$-wave production.
This corresponds to the quantum numbers of the $h_1$ axial vector meson.
In the present paper we study this reaction from the theoretical point of view. In the picture two $h_1$ states, corresponding to the $h_1(1170)$ and $h_1(1380)$,
together with their companions $f_1, a_1, b_1$ axial vector mesons, are dynamically generated from the pseudoscalar-vector meson interaction.
To produce $\eta' h_1(1380)$ we first produce $\eta'$ plus an extra pseudoscalar and a vector meson,
letting them propagate, using the chiral unitary approach to describe it, and the interaction generates the resonance.
Thus, we have a primary transition $J/\psi$ going to one vector and two pseudoscalars and one needs a theory for it.
Lacking a description for such a complicated dynamics, a symmetry is invoked, considering the $J/\psi$ to be an SU(3) flavor singlet.
The two reduced matrix elements needed are taken from the former study of the $J/\psi \to \omega \pi \pi$ and $J/\psi \to \phi \pi \pi$ reactions in Refs. \cite{ulfoller,Palomar}.
As we shall see, we are able to provide reasonable rates for this reaction
and for $J/\psi\rightarrow \pi^0b_1(1235)^0$,
and at the same time we make predictions for the rates of the $J/\psi\rightarrow\eta'h_1(1170)$ and $J/\psi\rightarrow\eta h_1(1170)$ $(\eta h_1(1380))$ reactions,
which constitute extra test for the dynamical origin of the axial vector mesons and we hope are measured in the near future.

\section{Formalism}
\label{sec:form}

\begin{figure}[b!]
\begin{center}
\includegraphics[scale=0.6]{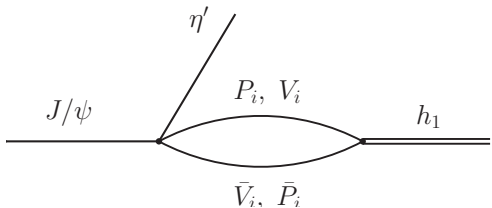}
\end{center}
\caption{Mechanism for $h_1$ resonance production, where $V_i \bar P_i$ are $K^* \bar K, \phi \eta, \omega \eta, \rho \pi$, and $h_1$ denotes $h_1(1170)$ or $h_1(1380)$.}
\label{Fig:1}
\end{figure}

In Ref. \cite{Luis} the $h_1(1170)$ and $h_1(1380)$ resonances were generated, albeit with somewhat different mass,
from the interaction of the channels $\frac{1}{\sqrt{2}}(K^* \bar K {-} \bar K^* K)$,
$\phi \eta, \omega \eta, \rho \pi$.
The $\eta' V$ ($\eta'$-vector) states have too large mass and were not included in Ref. \cite{Luis}.
We shall not take the theoretical masses of Ref. \cite{Luis}, but use the experimental ones to be accurate in the phase space calculations.
However, we shall take the couplings of the resonance to the different channels obtained in Ref. \cite{Luis} which are rather stable with respect to changes in the input of the model.
Given the nature assumed for these resonances as dynamically generated from the interaction of these channels,
the mechanism to produce them is depicted in Fig. \ref{Fig:1}.

The first thing we need is to describe the $J/\psi$ coupling to two pseudoscalars and one vector ($PPV$).
For this we rely on the work done in Ref. \cite{chic1},
where the experimental data on the $\chi_{c1} \to \eta \pi^+ \pi^-$ \cite{midhalo} were studied
and the $a_0(980)$ and $f_0(980)$ signals were well reproduced in the $\eta \pi$ and $\pi^+ \pi^-$ mass distributions, respectively.
The basic assumption done in Ref. \cite{chic1} was that $J/\psi$ was an SU(3) flavor singlet state and that it coupled to the structure Trace($PPP$).
Since this is not the only SU(3) singlet structure, in Ref. \cite{etac},
where the $\eta_c \to \eta \pi^+ \pi^-$ reaction was studied, it was shown that other possible structures lead to results in flagrant conflict with experiment.
Based on this finding we shall also take as the first step the structure
\begin{equation}\label{eq:LPPV}
  \mathcal{L}_{J/\psi, PPV}=A_1\; J/\psi_\mu \; {\rm Trace}(PPV^\mu),
\end{equation}
where $A_1$ is a constant, $P$ and $V$ are the
SU(3) $q\bar{q}$
matrices
written in terms of the pseudoscalar or vector mesons, respectively
\begin{equation}\label{eq:phimatrix}
P = \left(
           \begin{array}{ccc}
             \frac{1}{\sqrt{2}}\pi^0 + \frac{1}{\sqrt{3}}\eta + \frac{1}{\sqrt{6}}\eta' & \pi^+ & K^+ \\
             \pi^- & -\frac{1}{\sqrt{2}}\pi^0 + \frac{1}{\sqrt{3}}\eta + \frac{1}{\sqrt{6}}\eta' & K^0 \\
            K^- & \bar{K}^0 & -\frac{1}{\sqrt{3}}\eta + \sqrt{\frac{2}{3}}\eta' \\
           \end{array}
         \right),
\end{equation}
\begin{equation}\label{eq:Vmatrix}
V = \left(
           \begin{array}{ccc}
             \frac{1}{\sqrt{2}}\rho^0 + \frac{1}{\sqrt{2}}\omega  & \rho^+ & K^{*+} \\
             \rho^- & -\frac{1}{\sqrt{2}}\rho^0 + \frac{1}{\sqrt{2}}\omega  & K^{*0} \\
            K^{*-} & \bar{K}^{*0} & \phi \\
           \end{array}
         \right).
\end{equation}
In the pseudoscalar-meson matrix $P$, the mixing of $\eta$ and $\eta'$ is taken into account following Ref.~\cite{Bramon},
and the ideal mixing is assumed for $\omega$ and $\phi$ in the vector-meson matrix $V$.
From there we obtain
\begin{equation}\label{eq:TrPPV}
 {\rm Trace}(PPV)= (PPV)_{11}+(PPV)_{22}+(PPV)_{33},
\end{equation}
where
\begin{eqnarray}\label{eq:PPV11}
(PPV)_{11} & = & \left( \frac{\rho^0}{\sqrt{2}} + \frac{\omega}{\sqrt{2}} \right)\;
  \left[
    \left( \frac{\pi^0}{\sqrt{2}} + \frac{\eta}{\sqrt{3}} + \frac{\eta'}{\sqrt{6}} \right)^2 + \pi^+ \pi^- + K^+ K^-
  \right] \nonumber \\
&  & + \rho^- \left( \frac{2}{\sqrt{3}}\; \eta \pi^+ +  \frac{2}{\sqrt{6}} \;\eta' \pi^+ + K^+ \bar K^0 \right) \nonumber \\
&&+ K^{*-} \left( \frac{1}{\sqrt{2}} \; \pi^0 K^+ + \pi^+ K^0 + \sqrt{\frac{3}{2}} \; \eta' K^+ \right),
\end{eqnarray}
\begin{eqnarray}\label{eq:PPV22}
(PPV)_{22} & = & \rho^+ \left( \frac{2}{\sqrt{3}} \; \eta \pi^- +  \frac{2}{\sqrt{6}}\; \eta' \pi^- + K^0 K^- \right) \nonumber \\
&& {+}\left( \frac{-\rho^0}{\sqrt{2}} + \frac{\omega}{\sqrt{2}} \right)\;
  \left[ \pi^- \pi^+ +
    \left( \frac{-\pi^0}{\sqrt{2}} + \frac{\eta}{\sqrt{3}} + \frac{\eta'}{\sqrt{6}} \right)^2 + K^0 \bar K^0
  \right] \nonumber \\
&&+ \bar K^{*0} \left(  \pi^- K^+ -\frac{1}{\sqrt{2}} \; \pi^0 K^0 + \sqrt{\frac{3}{2}}\; \eta' K^0 \right),
\end{eqnarray}
\begin{eqnarray}\label{eq:PPV33}
(PPV)_{33} & = &  K^{*+} \left( \frac{1}{\sqrt{2}} \; K^- \pi^0  + \pi^- \bar K^0 + \sqrt{\frac{3}{2}} \; \eta' K^- \right) \nonumber \\
&& +  K^{*0} \left( K^- \pi^+  -\frac{1}{\sqrt{2}} \; \bar K^0 \pi^0  + \sqrt{\frac{3}{2}}\; \eta' \bar K^0 \right) \nonumber \\
&&+ \phi \left[ K^- K^+ + \bar K^0 K^0 + \left( -\frac{1}{\sqrt{3}}\eta + \sqrt{\frac{2}{3}}\eta' \right)^2 \right].
\end{eqnarray}
If we isolate the terms with $\eta'$, we obtain the coefficients $W_i$ for the different channels that we show in Table \ref{tab:wi1}.
\begin{table}[t]
     \renewcommand{\arraystretch}{1.2}
     \caption{Coefficients $W_i$ for channels involving $\eta'$ from the trace of $PPV$.}
\label{tab:wi1}
\centering
\begin{tabular}{c|ccccccccc}
 \hline\hline
   channel & $ \omega \eta \eta'$ & ~$\phi \eta \eta'$ & ~$K^{*+} K^- \eta'$ &~~$K^{*0} \bar K^0 \eta'$ & ~~$\bar K^{*0} K^0 \eta'$ & ~~$K^{*-}  K^+ \eta'$
   & ~~$\rho^0 \pi^0 \eta'$ & ~~$\rho^+ \pi^- \eta'$ & ~~$\rho^- \pi^+ \eta'$ \\
   \hline
   $W_i$  & $\frac{2}{3}$ & $-\frac{2\sqrt{2}}{3}$ & $\sqrt{\frac{3}{2}}$ & $\sqrt{\frac{3}{2}}$ & $\sqrt{\frac{3}{2}}$ & $\sqrt{\frac{3}{2}}$
   & $\sqrt{\frac{2}{3}}$  & $\sqrt{\frac{2}{3}}$  &  $\sqrt{\frac{2}{3}}$ \\
 \hline\hline
\end{tabular}
\end{table}

In Table \ref{tab:wi2} we show the coefficients $W_i$ for the terms {involving $\eta$ from} Trace($PPV$).
\begin{table}[t]
     \renewcommand{\arraystretch}{1.2}
     \caption{Coefficients $W_i$ for channels involving $\eta$ from the trace of $PPV$.}
\label{tab:wi2}
\centering
\begin{tabular}{c|cccccccc}
 \hline\hline
   channel & $ \omega \eta \eta$ & ~$\omega \eta \eta'$ & ~$\phi \eta \eta$ &~~$\phi \eta \eta'$ & ~~$ K^{*+} K^- \eta$ & ~~$K^{*0}  \bar K^0 \eta$
   & ~~$\bar K^{*0} K^0 \eta$ & ~~$K^{*-} K^+ \eta$  \\
   \hline
   $W_i$  & $\frac{\sqrt{2}}{3}$ & $\frac{2}{3}$ & $\frac{1}{3}$ & $-\frac{2\sqrt{2}}{3}$ & $0$ & $0$
   & $0$  & $0$   \\
 \hline\hline
  channel & $ ~\rho^0 \pi^0 \eta$ & ~~$ \rho^+ \pi^- \eta$ & ~~$ \rho^- \pi^+ \eta$ &~~ & ~~ & ~~
   & ~~& ~~  \\
   \hline
   $W_i$ & $\frac{2}{\sqrt{3}}$ & $\frac{2}{\sqrt{3}}$ & $\frac{2}{\sqrt{3}}$ &  &  &
   &   &    \\
 \hline\hline
 \end{tabular}
\end{table}

We shall also study the $J/\psi \to \pi^0 b_1(1235)^0$ reaction.
According to Ref. \cite{Luis} the $b_1(1235)$ couples to 
$\frac{1}{\sqrt{2}}\left( |K^* \bar K\rangle_{I=1} +|\bar K^* K \rangle_{I=1}\right)$,
$\phi \pi, \omega \pi, \rho \eta$
and the terms of the trace of $PPV$ where a $\pi^0$ appears have weights which we show in Table \ref{tab:wi3}.
\begin{table}[t!]
     \renewcommand{\arraystretch}{1.2}
     \caption{Coefficients $W_i$ for channels involving $\pi^0$ from the trace of $PPV$.}
\label{tab:wi3}
\centering
\begin{tabular}{c|ccccccc}
 \hline\hline
   channel & $ \pi^0 \phi \pi^0$ & ~$\pi^0 \omega \pi^0$ & ~$\pi^0 \rho^0 \eta$ &~~$\pi^0 K^{*+} K^-$ & ~~$ \pi^0 K^{*0} \bar K^0 $ & ~~$\pi^0 \bar K^{*0}  K^0$
   & ~~$\pi^0 K^{*-} K^+ $  \\
   \hline
   $W_i$  & $0$ & $\frac{1}{\sqrt{2}}$ & $\frac{2}{\sqrt{3}}$ & $\frac{1}{\sqrt{2}}$ & $-\frac{1}{\sqrt{2}}$ & $-\frac{1}{\sqrt{2}}$
   & $\frac{1}{\sqrt{2}}$    \\
 \hline\hline
\end{tabular}
\end{table}

It is interesting to see that with the coefficients $W_i$ of Table \ref{tab:wi1} for $\eta' K^* \bar K$ or $\eta' \bar K^* K$,
we find the same coupling for the terms of the combination
\begin{equation}\label{eq:combKsK0}
{(K^* \bar K)_{I=0} \equiv \frac{1}{\sqrt{2}}\left(|K^* \bar K\rangle_{I=0} -|\bar K^* K\rangle_{I=0}\right)} =-\frac{1}{2} (K^{*+} K^- + K^{*0} \bar K^0 +K^{*-} K^+ + \bar K^{*0} K^0),
\end{equation}
which is the $K^* \bar K,
\bar K^* K$ combination
with isospin $I=0$ and $C$-parity negative that we obtain ($K^- =- | 1/2, -1/2\rangle, K^{*-}=-| 1/2, -1/2\rangle, C K^{*+}= -K^{*-}$),
as it corresponds to the $h_1$ state.
We also find a common coupling to this state in Table \ref{tab:wi2} for channels with $\eta$, but the $W_i$ coefficient is zero.
Similarly, in Tables \ref{tab:wi1} and \ref{tab:wi2} we also see a common coupling to the terms of the combination
\begin{equation}\label{eq:combrhopi}
 |\rho \pi \rangle_{I=0} =-\frac{1}{\sqrt{3}} (\rho^+ \pi^- + \rho^- \pi^+ + \rho^0 \pi^0),
\end{equation}
which is our $\rho \pi$ state with $I=0$ ($\rho^+ = -| 1,1 \rangle, \pi^+ = - |1,1 \rangle$).

For the terms that contain a $\pi^0$ in Table \ref{tab:wi3} we find a common coupling to the terms of the combination
\begin{equation}\label{eq:combKsK1}
 {(K^* \bar K)_{I=1} \equiv \frac{1}{\sqrt{2}}\left( |K^* \bar K\rangle_{I=1} +|\bar K^* K \rangle_{I=1}\right)}
 =-\frac{1}{2} (K^{*+} K^- - K^{*0} \bar K^0 +K^{*-} K^+ - \bar K^{*0} K^0),
\end{equation}
which has $I=1$ and $C$-parity negative, as it corresponds to the $b_1$ state.

In Tables \ref{tab:gh1} and \ref{tab:gb1} we show the couplings of the resonances $h_1(1170), h_1(1380)$ and $b_1(1235)$ to the different channels,
which are taken from Ref. \cite{Luis}\footnote{Note that the couplings of Ref.~\cite{Luis} are calculated for $\frac{1}{\sqrt{2}}\left(|\bar{K}^* K\rangle_{I=0} -|K^* \bar{K}\rangle_{I=0}\right)$ {for $h_1$ and $\frac{1}{\sqrt{2}}\left(|\bar{K}^* K\rangle_{I=1} +|K^* \bar{K}\rangle_{I=1}\right)$ for $b_1$} while here we use $\frac{1}{\sqrt{2}}\left(|K^* \bar K\rangle_{I=0} -|\bar K^* K\rangle_{I=0}\right)$ for $h_1$ states, and $\frac{1}{\sqrt{2}}\left(|K^* \bar K\rangle_{I=1} +|\bar K^* K\rangle_{I=1}\right)$ for the $b_1$.
As a consequence, the couplings of $h_1$ to $(K^*\bar{K})_{I=0}$ in Table~\ref{tab:gh1} have opposite sign to those in Ref.~\cite{Luis} while {the coupling of $b_1$ to $(K^*\bar{K})_{I=1}$ in} Table~\ref{tab:gb1} has the same sign.}.
\begin{table}[t!]
     \renewcommand{\arraystretch}{1.2}
     \caption{Couplings $g_{R,i}$ of the resonance $h_1(1170)$ or $h_1(1380)$ to the different channels {in units of $\mev$}.}
\label{tab:gh1}
\centering
\begin{tabular}{c|cccc}
 \hline\hline
    & ${ (K^* \bar K )_{I=0}}$ & $\phi \eta$ &  $\omega \eta$ &  $\rho \pi$ \\
   \hline
   $h_1(1170)$  &{ $-781+i498$} &  $46 -i13$    &  $23 - i28$ & ~$-3453+i1681$~ \\
   $h_1(1380)$ & {$-6147-i183$} &  ~$-3311 +i47$~ &  ~$3020-i22$~ & $648 - i959$   \\
 \hline\hline
\end{tabular}
\end{table}
\begin{table}[t!]
     \renewcommand{\arraystretch}{1.2}
     \caption{Couplings $g_{R,i}$ of the resonance $b_1(1235)$ to the different channels {in units of $\mev$}.}
\label{tab:gb1}
\centering
\begin{tabular}{c|cccc}
 \hline\hline
               & ${(K^* \bar K )_{I=1}}$  &  $\phi \pi$    &   $\omega \pi$    &  $\rho \eta$   \\
   \hline
   $b_1(1235)$ & ~{{$6172-i75$}}~                &~$2087 - i385$~ &  ~$-1869+i300$~   &  ~${-}3041+i498$~   \\
 \hline\hline
\end{tabular}
\end{table}

As one can see in Table~\ref{tab:gh1}, the couplings of $h_1(1170)$ to the $\phi\eta$ and $\omega\eta$ channels are small.
Then, we omit the contribution from these channels in the calculation of $J/\psi\rightarrow\eta'(\eta)h_1(1170)$.

In view of Eqs. \eqref{eq:combKsK0}, \eqref{eq:combrhopi} and \eqref{eq:combKsK1}, we can write the coupling of the resonance to an individual channel as
\begin{eqnarray}\label{eq:gRi}
g_{R, K^{*+} K^-} &=&- \frac{1}{2} \ g_{R, (K^* \bar K)_{I=0}}, ~ R= h_1(1170), h_1(1380) \nonumber \\
g_{R, K^{*+} K^-} &=&-  \frac{1}{2} \ g_{R, (K^* \bar K)_{I=1}}, ~ R= b_1(1235)  \\
g_{R, \rho^+ \pi^-}&=&-  \frac{1}{\sqrt{3}} \ g_{R, (\rho \pi)_{I=0}}, ~  R= h_1(1170), h_1(1380).\nonumber
\end{eqnarray}
Then, apart from a factor $A_1 \, \vec \epsilon_{J/\psi} \cdot \vec \epsilon_V $, the transition matrices are given by
\begin{eqnarray}\label{eq:th1etap}
t_{J/\psi, \;\eta' h_i} &=& -2 \; W_{K^{*+} K^- \eta'} \; G_{K^* \bar K} \; g_{R, (K^* \bar K)_{I=0}}
+ W_{\phi \eta \eta'} \; G_{\phi \eta} \; g_{R, \phi \eta} \nonumber \\
&& + W_{\omega \eta \eta'} \; G_{\omega \eta} \; g_{R, \omega \eta}
- \sqrt{3} \; W_{\rho^+ \pi^- \eta'} \; G_{\rho \pi} \; g_{R, (\rho \pi)_{I=0}},
\end{eqnarray}
\begin{eqnarray}\label{eq:th1eta}
t_{J/\psi, \;\eta h_i} &=& 2 \; W_{\eta \omega \eta} \; G_{\omega \eta} \; g_{R, \omega \eta}
+ 2\; W_{\eta \phi \eta} \; G_{\phi \eta} \; g_{R, \phi \eta} \nonumber \\
&& - \sqrt{3} \; W_{\eta \rho^+ \pi^- } \; G_{\rho \pi} \; g_{R, (\rho \pi)_{I=0}},
\end{eqnarray}
for $h_i$, either of the two $h_1$ states,
where the extra factor two in the $\omega \eta$ and $\phi \eta$ channels {in Eq. \eqref{eq:th1eta}} comes from the identity of the two $\eta$ particles in the channels.
Finally,
\begin{eqnarray}\label{eq:tb1}
t_{J/\psi, \;\pi^0 b_1} &=& -2 \; W_{\pi^0 K^{*+} K^-} \; G_{K^* \bar K} \; g_{R, (K^* \bar K)_{I=1}}
 {+ 2\; W_{\pi^0 \phi \pi^0} \; G_{\phi \pi} \; g_{R, \phi \pi}}\nonumber \\
&& + 2\; W_{\pi^0 \omega \pi^0} \; G_{\omega \pi} \; g_{R, \omega \pi} +  W_{\pi^0 \rho^0 \eta} \; G_{\rho \eta} \; g_{R, \rho \eta},
\end{eqnarray}
with $R$ standing for $b_1$.
{Prefactors $2$ in front of $W_{\pi^0\phi\pi^0}$ and $W_{\pi^0\omega\pi^0}$ appear from the identity of two $\pi^0$.}

In Eqs.~\eqref{eq:th1etap}, \eqref{eq:th1eta}, and \eqref{eq:tb1},
$G_i$, with $i$ being $K^*\bar{K},\phi\eta,\omega\eta,\rho\pi,\phi\pi,\omega\pi,\rho\eta$, is the meson-meson loop function.
Here, $G_i$ is regularized with dimensional regularization as in Ref.~\cite{Luis}, and for $G_i$ with $K^*$ or $\rho$ the vector meson mass is smeared with the spectral function (see $e.g.$ Ref.~\cite{rocageng} for details){, and the $K^*\bar{K}$ loop function is averaged over the isospin after the convolution; $G_{K^*\bar{K}}=(G_{K^{*+}K^-}+G_{K^{*0}\bar{K}^0})/2$.}

{With the amplitudes Eqs.~\eqref{eq:th1etap}, \eqref{eq:th1eta}, and \eqref{eq:tb1}, the decay width of $J/\psi\rightarrow PR$ ($PR=\eta^{(\prime)}h_1(1170),\eta^{(\prime)}h_1(1380),$ and $\pi^0b_1(1235)^0$) is obtained by
\begin{align}
 \Gamma_{J/\psi\rightarrow PR}=\frac{p_P}{8\pi M_{J/\psi}^2}|t_{J/\psi,PR}|^2
\end{align}
with $p_P=\lambda^{1/2}(M_{J/\psi}^2,m_P^2,M_R^2)/[2M_{J/\psi}]$ {and} ${\lambda(x,y,z)=x^2+y^2+z^2-2xy-2yz-2zx}$.}

\section{Connection with the $\boldsymbol{J/\psi\rightarrow\omega\pi\pi,\phi\pi\pi}$ reactions}
\label{sec:con}
One attractive thing of the ${J/\psi PPV}$ Lagrangian considered is that it automatically fulfills the Okubo-Zweig-Iizuka (OZI) rule.
As we can see in $(PPV)_{33}$, Eq.~(\ref{eq:PPV33}), the $\phi\pi\pi$ channel does not appear.
The $\phi$, implicitly assumed there as a $s\bar{s}$ state, only couples to kaons or $\eta$, $\eta'$ that contain $s$ quarks.
Actually the $J/\psi\rightarrow\phi\pi\pi$ decay is suppressed with respect to $\omega\pi\pi$ by one order of magnitude,
but it is not zero.
In Refs.~\cite{ulfoller,Palomar} an extra Lagrangian was used, that in our new formalism can be cast as
\begin{align}
 \mathcal{L}'_{J/\psi,PPV}=&\beta A_1\; (J/\psi)_\mu \; \tr(PP)\; \tr(V^\mu), \label{eq:lprime}
\end{align}
where
\begin{align}
 \tr(PP)=&(PP)_{11}+(PP)_{22}+(PP)_{33},
\end{align}
with
\begin{align}
 (PP)_{11}=&\left(\frac{\pi^0}{\sqrt{2}}+\frac{\eta}{\sqrt{3}}+\frac{\eta'}{\sqrt{6}}\right)^2+\pi^+\pi^-+K^+K^-,\\
 (PP)_{22}=&\pi^-\pi^++\left(-\frac{\pi^0}{\sqrt{2}}+\frac{\eta}{\sqrt{3}}+\frac{\eta'}{\sqrt{6}}\right)^2+K^0\bar{K}^0, \\
 (PP)_{33}=&K^-K^++\bar{K}^0K^0+\left(-\frac{\eta}{\sqrt{3}}+\sqrt{\frac{2}{3}}\eta'\right)^2,\\
 \tr(V)=&\sqrt{2}\omega+\phi.
\end{align}
We shall call $W'_i$ the weights for the different channels discussed above.
In Table~\ref{tab:vi}, we show the weights $W'_i$ which appear in our calculation.
To include these new terms we only need minor changes:
\begin{table}[t]
     \renewcommand{\arraystretch}{1.2}
     \caption{Weights $W'_i$ appearing in ${\rm Trace}(PP)\; {\rm Trace}(V)$.}
\label{tab:vi}
\centering
\begin{tabular}{c|cccc}
 \hline\hline
 &$\phi\eta\eta$ &~$\omega\eta\eta$~ &~$\pi^0\phi\pi^0$~ &~$\pi^0\omega\pi^0$~ \\\hline
  $W'_i$&$1$ &$\sqrt{2}$ &$1$ &$\sqrt{2}$ \\
 \hline\hline
\end{tabular}
\end{table}

\begin{enumerate}
\item[i)] Changes in Eq.~(\ref{eq:th1etap}): \\
  \begin{align}
	  W_{\eta\phi\eta}\rightarrow W_{\eta\phi\eta}+\beta W'_{\eta\phi\eta}, \label{eq:13-1a}\\
	  W_{\eta\omega\eta}\rightarrow W_{\eta\omega\eta}+\beta W'_{\eta\omega\eta}.
	 \end{align}
\item[ii)] Changes in Eq.~(\ref{eq:th1eta}): \\
	 \begin{align}
	  W_{\pi^0\phi\pi^0}\rightarrow W_{\pi^0\phi\pi^0}+\beta W'_{\pi^0\phi\pi^0},\label{eq:13-1b}\\
	  W_{\pi^0\omega\pi^0}\rightarrow W_{\pi^0\omega\pi^0}+\beta W'_{\pi^0\omega\pi^0}.
	 \end{align}
\end{enumerate}

It is easy to establish the correspondence of $A_1, \beta$ to the parameters $\tilde{g}$ and $\nu$ of Ref.~\cite{Palomar}
or the $\lambda_\phi$ parameter of Ref.~\cite{ulfoller}.
For this we choose the $J/\psi\rightarrow\omega\pi^+\pi^-$, $J/\psi\rightarrow\phi\pi^+\pi^-$, $J/\psi\rightarrow\omega K^+K^-$, $J/\psi\rightarrow\phi K^+K^-$ transitions and take the weights from the Lagrangian of Eqs.~(\ref{eq:LPPV}) and (\ref{eq:lprime}) and compare them with the results of Ref.~\cite{Palomar} (see in Eq.~(18) of that work the term without rescattering, the term $1$ in $(1+Gt)$).
We show these terms in Table~\ref{tab:vii}, from where
\begin{align}
 A_1\equiv -\tilde{g};~~~ \beta\equiv\frac{\nu-1}{3}.
\end{align}
\begin{table}[t]
\caption{Amplitudes for $J/\psi\rightarrow\phi(\omega)\pi^+\pi^-(K^+K^-)$.}
 \label{tab:vii}
 \centering
 \begin{tabular}{c|cc}
  \hline\hline
  &Ref.~\cite{Palomar} &Present work \\\hline
  $\omega\pi^+\pi^-$&$~-\tilde{g}\frac{\sqrt{2}}{3}(1+2\nu)$ ~~&~~$A_1\sqrt{2}+A_1\beta 2\sqrt{2}$~ \\
  $\phi\pi^+\pi^-$&$-\tilde{g}\frac{2}{3}(\nu-1)$ &$0+A_1\beta \sqrt{2}$ \\
  $\omega K^+K^-$&$-\tilde{g}\frac{1}{3\sqrt{2}}(4\nu-1)$ &$A_1\frac{1}{\sqrt{2}}+A_1\beta\sqrt{2}$ \\
  $\phi K^+K^-$&$-\tilde{g}\frac{1}{3}(1+2\nu)$ &$A_1+A_1\beta$ \\
  \hline\hline
 \end{tabular}
 \end{table}
In addition, in Ref.~\cite{Palomar} the relationship of $\nu$ to the parameter $\lambda_\phi$ of Ref.~\cite{ulfoller} was found as
\begin{align}
 \nu=\frac{\sqrt{2}+2\lambda_\phi}{\sqrt{2}-\lambda_\phi}.
\end{align}

Note that $\nu=1$ corresponds to $\beta=0$ and is the case with the OZI respecting Lagrangian.
In Ref.~\cite{Palomar} two solutions were found depending on the sign of the anomalous $J/\psi\rightarrow\rho\pi$ terms used in that work:
 \begin{enumerate}
  \item[i)] \hfill \makebox[0pt][r]{%
            \begin{minipage}[b]{\textwidth}
              \begin{equation}
	       \tilde{g}=0.032\pm0.001;~~~ \lambda_\phi=0.12\pm0.03 ~~ (\beta=0.0927)\label{eq:14-1a}
              \end{equation}
          \end{minipage}}
  \item[ii)]	\hfill \makebox[0pt][r]{%
            \begin{minipage}[b]{\textwidth}
              \begin{equation}
	       \tilde{g}=0.015\pm0.001;~~~ \lambda_\phi=0.20\pm0.03 ~~ (\beta=0.165)\label{eq:14-1b}
             \end{equation}
          \end{minipage}}
 \end{enumerate}
The results for $\lambda_\phi$ in Ref.~\cite{ulfoller} are
\begin{align}
 \lambda_\phi=0.17\pm 0.06.
\end{align}
We shall evaluate the rates with the two sets of values of Eqs.~(\ref{eq:14-1a}) and (\ref{eq:14-1b}).

\section{Results}
\label{sec:res}
In Table~\ref{tab:viii} we show the branching ratios that we obtain for the different reactions.
\begin{table}[t]
\caption{Branching fractions (a) and (b) with Eq.~(\ref{eq:14-1a}) and Eq.~(\ref{eq:14-1b}) for the parameter set of $\tilde{g}$ and $\beta$, respectively.}
 \label{tab:viii}
 \centering
 \begin{tabular}[t]{c|cc}
  \hline\hline
  &branching fraction (a)~~&~~branching fraction (b)\\\hline
  $\br[J/\psi\rightarrow\eta'h_1(1380)]$& ${2.35}\tento{-3}$ & ${5.16}\tento{-4}$\\
  $\br[J/\psi\rightarrow\eta h_1(1380)]$& $3.65\tento{-5}$ & $1.02\tento{-5}$\\
  $\br[J/\psi\rightarrow\eta' h_1(1170)]$& ${5.35}\tento{-4}$ & ${1.18}\tento{-4}$\\
  $\br[J/\psi\rightarrow\eta h_1(1170)]$& $9.49\tento{-4}$ &$2.08\tento{-4}$\\
  $\br[J/\psi\rightarrow\pi^0b_1(1235)^0]$& $1.23\tento{-3}$ & $2.77\tento{-4}$\\
  \hline\hline
 \end{tabular}
 \end{table}
The results for $J/\psi\rightarrow\pi^0b_1(1235)^0$ can be compared with those of the PDG \cite{PDG}:
\begin{align}
 \br[J/\psi\rightarrow \pi^0b_1(1235)^0]=&(2.3\pm0.6)\tento{-3},\\
 \br[J/\psi\rightarrow \pi^\pm b_1(1235)^\mp]=&(3.0\pm0.5)\tento{-3}.
\end{align}
Our result for $J/\psi\rightarrow\pi^0b_1(1235)^0$ branching fraction,
$1.23\tento{-3}$ of version (a) in Table~\ref{tab:viii}, with about 10\% error, is compatible with $\frac{1}{3}\br[J/\psi\rightarrow(\pi^\pm b_1(1235)^\mp+\pi^0b_1(1235)^0)]$ that we would obtain assuming isospin symmetry, $(1.77\pm 0.4)\tento{-3}$, from Ref.~\cite{PDG}.
The comparison of the results for the $J/\psi\rightarrow \eta'h_1(1380)$ with the experiment of Ref.~\cite{Besnew} requires extra work that we conduct in the next section.
Apart from the absolute values of the branching ratios, the ratios of the rates between different decays are very illustrative.
In Table~\ref{tab:ix}, we show the ratios that we obtain.
\begin{table}[t]
\caption{Ratios to the $J/\psi\rightarrow\pi^0b_1(1235)^0$ width.
 The meaning of the cases (a) and (b) is the same as Table~\ref{tab:viii}.}
 \label{tab:ix}
 \centering
 \begin{tabular}[t]{c|cc}
  \hline\hline
  &~ratio (a)~~&~~ratio (b)~\\\hline
  $\br[J/\psi\rightarrow\eta'h_1(1380)]/\br[J/\psi\rightarrow\pi^0 b_1(1235)^0]$& ~${1.91}$~ &~${1.86}$~\\
  $\br[J/\psi\rightarrow\eta h_1(1380)]/\br[J/\psi\rightarrow\pi^0 b_1(1235)^0]$& ~$0.03$~ &~$0.04$~\\
  $\br[J/\psi\rightarrow\eta' h_1(1170)]/\br[J/\psi\rightarrow\pi^0 b_1(1235)^0]$ &${0.44}$& ${0.43}$ \\
  $\br[J/\psi\rightarrow\eta h_1(1170)]/\br[J/\psi\rightarrow\pi^0 b_1(1235)^0]$&$0.77$ & $0.75$ \\
  \hline\hline
 \end{tabular}
 \end{table}
As we can see, the ratios are rather independent on which option (a) of or (b) of Eqs.~(\ref{eq:14-1a}) and (\ref{eq:14-1b}) we take.
It is remarkable that while these ratios are of the order of unity,
the ratio $\br[J/\psi\rightarrow\eta h_1(1380)]/\br[J/\psi\rightarrow\pi^0b_1(1235)^0]$ is of the order of $3\tento{-2}$.
This could indicate that it is just more than an accident that the $J/\psi\rightarrow\eta'h_1(1380)$ rate has been observed but not the related one $J/\psi\rightarrow\eta h_1(1380)$.

\section{Evaluation of the case of $\boldsymbol{\br(J/\psi\rightarrow\eta'h_1(1380))\times\br(h_1(1380)\rightarrow(K^{*+}K^-+c.c.))}$ counted in the $\boldsymbol{\eta'K^+K^-\pi^0}$ mode}
In Ref.~\cite{Besnew} the branching ratio of $J/\psi\rightarrow \eta'h_1(1380)$ times the branching ratio $\br(h_1(1380)\rightarrow K^{*+}K^-+c.c.)$ is measured in the $\eta'K^+K^-\pi^0$ mode.
A branching ratio for this fraction of $(1.51\pm0.09\pm0.21)\tento{-4}$ is obtained while a $\br(J/\psi\rightarrow\eta'h_1(1380))\times\br(h_1(1380)\rightarrow K^*\bar{K}+c.c.)=(2.16\pm 0.12\pm 0.29)\tento{-4}$ is observed in the $\eta'K_S^0K^\pm\pi^\mp$ mode.
Another interesting finding in that work is that the mass of the $h_1(1380)$ is better fitted with $M=(1441.7\pm 4.9)~\mev$, bigger than the nominal one of the PDG, $M=(1407\pm 12)~\mev$.
Let us see how we approach these issues.

In Table~\ref{tab:x}, we show the different channels in $K^*\bar{K}+c.c.$ and their decay modes,
together with the weight of each mode given by the square of the corresponding isospin Clebsch-Gordan (CG) coefficient.
\begin{table}[t]
 \caption{The decay modes of $K^*\bar{K}+c.c.$.
 In parenthesis the weight of this channel.}
 \label{tab:x}
 \centering
 \begin{tabular}[t]{c|cccc}
  \hline\hline
  channel&$K^{*+}K^-$ &$K^{*-}K^+$ &$K^{*0}\bar{K}^0$ &$\bar{K}^{*0}K^0$ \\\hline
  &~$K^+\pi^0K^-(1/3)$~~ &~~$K^-\pi^0K^+(1/3)$~~ &~~$K^0\pi^0\bar{K}^0(1/3)$~~ &~~$\bar{K}^0\pi^0K^0(1/3)$~~ \\
  &$K^0\pi^+K^-(2/3)$ &$\bar{K}^0\pi^+K^-(2/3)$ &$K^+\pi^-\bar{K}^0(2/3)$ &$K^-\pi^+K^0(2/3)$ \\
  \hline\hline
 \end{tabular}
\end{table}
In $K^{*+}K^-+K^{*-}K^+$ measured in $K^+K^-\pi^0$ mode the weight is $\frac{1}{2}\cdot\frac{1}{3}$ of the weight for the decay in all channels.
In $K^*\bar{K}+c.c.$ measured in the $K_S^0K^\pm\pi^\mp$ one counts $\frac{1}{2}\cdot \frac{2}{3}$ of the sum of the rates since the $K_S^0K^\pm\pi^\mp$ is selected in the $K^{*\pm}$ mode and there is
an extra
factor $1/2$ reduction for measuring $K_S^0$ rather than $K^0$ or $\bar{K}^0$.
Thus the two rates should be equivalent assuming isospin symmetry as indicated in Ref.~\cite{Besnew}.
Actually, within errors the two experimental rates are compatible, although some isospin violation is claimed in Ref.~\cite{Besnew}.
We will stick to the isospin symmetry for the vertices,
but some isospin violation necessarily appears in the final results due to differences in the phase space due to different masses of the particles.
{However, in the present study we do not enter the issue of the isospin violation, and focus only on the $K^+K^-\pi^0$ mode.}

\begin{figure}[t]
 \centering
 \includegraphics[width=8cm]{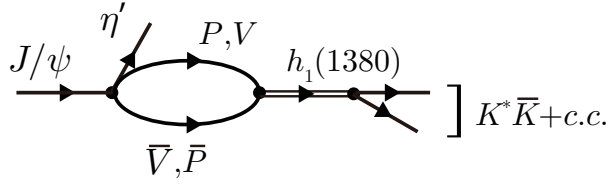}
 \caption{Diagram for the $J/\psi\rightarrow\eta'K^*\bar{K}+c.c.$ through $h_1(1380)$.}
 \label{fig:2}
\end{figure}

The exercise done above indicates that we get a reduction of a factor of $6$
from the rate obtained for $J/\psi\rightarrow \eta'h_1(1380)$ in Table~\ref{tab:viii},
assuming it to be dominated by the $K^*\bar{K}+c.c.$ channel.
This makes the new rate much closer to experiment, but still larger by about a factor $2$-$2.5$.
We then conduct the investigation forward and perform two different evaluations:

1) $J/\psi\rightarrow\eta'K^*\bar{K}+c.c.$:\\
Here we consider the $K^*\bar{K}$ in the final state explicitly as shown in Fig.~\ref{fig:2}.
The differential width for this channel is given by
\begin{align}
 \frac{d\Gamma}{d\minv(K^*\bar{K})}=\frac{1}{(2\pi)^3}\frac{1}{4M_{J/\psi}^2}p_{\eta'}\tilde{p}_{\bar{K}}\overline{\sum}\sum|t|^2\label{eq:18-1}
\end{align}
with
\begin{align}
 p_{\eta'}=&\frac{1}{2M_{J/\psi}}\lambda^{1/2}(M_{J/\psi}^2,m_{\eta'}^2,\minv^2(K^*\bar{K})),\\
 \tilde{p}_{\bar{K}}=&\frac{1}{2\minv(K^*\bar{K})}\lambda^{1/2}(\minv^2(K^*\bar{K}),M_{K^*}^2,m_{\bar{K}}^2),
\end{align}
where we
will take
the masses for $K^{*+}K^-$.

The $t$ matrix in Eq.~(\ref{eq:18-1}) is given by
\begin{align}
 t=\tilde{t}\,\vec{\epsilon}_{J/\psi}\cdot\vec{\epsilon}_{K^{*}}\label{eq:19-1}
\end{align}
with
\begin{align}
 \tilde{t}=&A_1t_{J/\psi,\eta'h_1}\frac{1}{\minv^2(K^*\bar{K})-M_{h_1}^2+iM_{h_1}\Gamma_{h_1}}\cdot g_{R,(K^*\bar{K})_{I=0}}\label{eq:19-2}
\end{align}
with $t_{J/\psi,\eta'h_1}$ given by Eq.~(\ref{eq:th1etap}) together with Eq.~(\ref{eq:13-1a}) and $g_{R,(K^*\bar{K})_{I=0}}$ taken from Table~\ref{tab:gh1}.
We also take the mass and width of the $h_1(1380)$ from the PDG \cite{PDG}, $M_{h_1}=1407~\mev$ and $\Gamma_{h_1}=89~\mev$.
By taking the coupling $g_{R,(K^*\bar{K})_{I=0}}$ we are automatically summing all four final $K^*\bar{K}+\bar{K}^*K$ channels.
The $\vec{\epsilon}_{J/\psi}\cdot\vec{\epsilon}_{K^*}$ coupling stems from the primary vertex $\vec{\epsilon}_{J/\psi}\cdot\vec{\epsilon}_{V}$
together with the $\vec{\epsilon}_{V}\cdot\vec{\epsilon}_{h_1}$ from the $h_1\rightarrow PV$ vertex \cite{Luis},
and the final $\vec{\epsilon}_{h_1}\cdot\vec{\epsilon}_{K^*}$ vertex, after summing over the polarizations of the intermediate $V$ and $h_1$.
The sum and  average over polarizations of $\vec{\epsilon}_{J/\psi}\cdot\vec{\epsilon}_{K^*}$ in Eq.~(\ref{eq:19-1}) gives unity.

Integrating Eq.~(\ref{eq:18-1}) over the $K^*\bar{K}$ invariant mass we should get a factor 6 times bigger rate than the one
obtained projecting over the measured channels in Ref.~\cite{Besnew}.
Hence, we should divide the
results obtained by a factor 6 for comparison with the experimental numbers.

\begin{figure}[t]
 \centering
 \includegraphics[width=8cm]{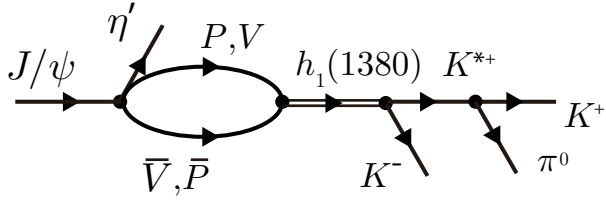}
 \caption{Diagram for the $J/\psi\rightarrow\eta'K+K^-\pi^0$ through $h_1(1380)$ and $K^{*+}$.}
 \label{fig:3}
\end{figure}

2) $J/\psi\rightarrow\eta'h_1(1380),h_1(1380)\rightarrow K^{*+}K^-,K^{*+}\rightarrow K^+\pi^0$:\\
We take into account this channel by looking at the mechanism of Fig.~\ref{fig:3}.
Here we take into account explicitly the $h_1(1380)$ and $K^{*+}$ propagators,
and hence, automatically the mass distribution of these two states.
We have now the double differential mass distribution \cite{Sakai}
\begin{align}
 \frac{d^2\Gamma}{d\minv(K^{*+}K^-)d\minv(K^+\pi^0)}=\frac{1}{(2\pi)^5}\;\frac{1}{4M_{J/\psi}^2}\;p_{\eta'}\;p_{K^-}\;\tilde{p}_{K^+}\;|t'|^2,
\end{align}
where
\begin{align}
 p_{\eta'}=&\frac{1}{2M_{J/\psi}}\; \lambda^{1/2}(M_{J/\psi}^2,m_{\eta'}^2,\minv^2(K^{*+}K^-)),\\
 p_{K^-}=&\frac{1}{2\minv(K^{*+}K^-)}\; \lambda^{1/2}(\minv^2(K^{*+}K^-),m_{K^-}^2,\minv^2(K^+\pi^0)),\\
 \tilde{p}_{K^+}=&\frac{1}{2\minv(K^+\pi^0)}\; \lambda^{1/2}(\minv^2(K^+\pi^0),m_{K^+}^2,m_{\pi^0}^2),\label{eq:20-1}
\end{align}
and $t'$ is now given by
\begin{align}
 t'=\frac{1}{2}\tilde{t}\; D_{K^*}(\minv(K^+\pi^0))\; \frac{1}{\sqrt{3}}\; g_{K^*,K\pi}\; \vec{\epsilon}_{J/\psi}\cdot\vec{\tilde{p}}_K,\label{eq:21-1}
\end{align}
where {$\tilde{t}$ is given by Eq.~(\ref{eq:19-2})} and the effective coupling $g_{K^*,K\pi}$ is defined below calculating the $K^*$ width,
and
\begin{align}
 D_{K^*}(\minv(K^+\pi^0))=&\frac{1}{\minv^2(K^+\pi^0)-M_{K^*}^2+iM_{K^*}\Gamma_{K^*}}
\end{align}
where, once again, the $\vec{\epsilon}_{J/\psi}\cdot\vec{\tilde{p}}_K$ factor appears after summing over the
polarization of the {internal} vector
{and} axial vector mesons.
The factor $1/2$ accounts for the coupling of the resonance to $K^{*+}K^-$, $-g_{R,(K^*\bar{K})_{I=0}}/2$,
and the $(-)\frac{1}{\sqrt{3}}$ the CG coefficient of $K^{*+}\rightarrow K^+\pi^0$.
Taking into account that the full width of the $K^*$ is given by
\begin{align}
 \Gamma_{K^*}=\frac{1}{8\pi}\frac{1}{M_{K^*}^2}\tilde{p}_{K^+}g_{K^*,K\pi}^2\frac{1}{3}\tilde{p}_{K^+}^2
\end{align}
and that $\overline{\sum}\left|\vec{\epsilon}_{J/\psi}\cdot\vec{\tilde{p}}_K\right|^2=\tilde{p}_K^2/3$,
we obtain
\begin{align}
 \frac{d^2\Gamma}{d\minv(K^{*+}K^-)d\minv(K^+\pi^0)}=&\frac{1}{(2\pi)^5}p_{\eta'}p_{K^-}\frac{1}{4M_{J/\psi}^2}\frac{1}{3}8\pi M_{K^*}^2\Gamma_{K^*}\frac{1}{4}|\tilde{t}|^2|D_{K^*}(\minv(K^+\pi^0))|^2\label{eq:21-2}
\end{align}
with $\tilde{p}_{K^+}$ given by Eq.~(\ref{eq:20-1}), and
\begin{align}
 \Gamma_{K^*}=&{\Gamma_{K^*}^0}\left(\frac{\tilde{p}_{K^+}}{\tilde{p}_{0,K^+}}\right)^3;~~~{\Gamma_{K^*}^0}={50}~\mev,\\
 \tilde{p}_{0,K^+}=&\frac{1}{2M_{K^*}}\lambda^{1/2}(M_{K^*}^2,m_{\pi^0}^2,m_{K^+}^2).
\end{align}
Finally, let us recall that we have calculated $J/\psi\rightarrow\eta'K^{*+}K^-$ and then projected to $K^+K^-\pi^0$,
but to match the experiment we have to add the $J/\psi \to \eta'K^{*-}K^+$ projected to $K^+K^-\pi^0$, which gives the same contribution.
Thus, we have to multiply by a factor two the results obtained by means of Eq.~(\ref{eq:21-2}).

\section{Results for $\boldsymbol{d^2\Gamma/d\minv(K^+K^-\pi^0)d\minv(K^+\pi^0)}$}
\begin{figure}[t]
 \centering
 \includegraphics[width=12cm]{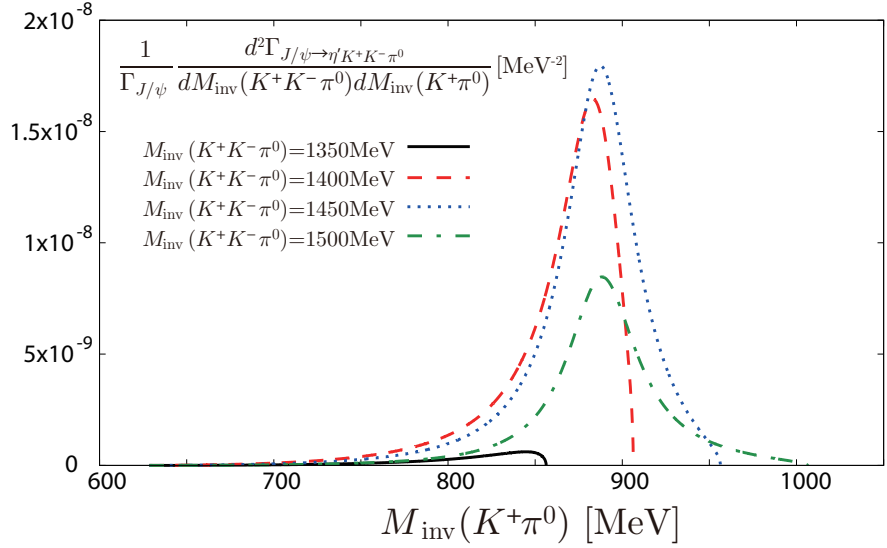}
 \caption{$[d^2\Gamma/d\minv(K^+K^-\pi^0)d\minv(K^+\pi^0)]/\Gamma_{J/\psi}$ as a function of $\minv(K^+\pi^0)${, fixing $\minv(K^+K^-\pi^0)$ as 1350, 1400, 1450 and 1500 MeV, respectively}.}
 \label{fig:4}
\end{figure}

In Fig.~\ref{fig:4}, we show the results for $d^2\Gamma/d\minv(K^+K^-\pi^0)d\minv(K^+\pi^0)$ as a function of $\minv(K^+\pi^0)$ for different values of $\minv(K^+K^-\pi^0)$
around the nominal mass of the $h_1(1380)$, $1350~\mev$, $1400~\mev$, $1450~\mev$, $1500~\mev$.
{Here, we only show the results with the parameter set (a) given in Eq.~\eqref{eq:14-1a}.}
We can see the shape of $K^*$ resonance in the figure for all
the $K^+K^-\pi^0$ invariant masses except for the lowest one of $1350~\mev$.
Certainly, for energies above the $K^*\bar{K}$ threshold, ${1385}~\mev$, the shape is better reproduced.

\begin{figure}[t]
 \centering
 \includegraphics[width=12cm]{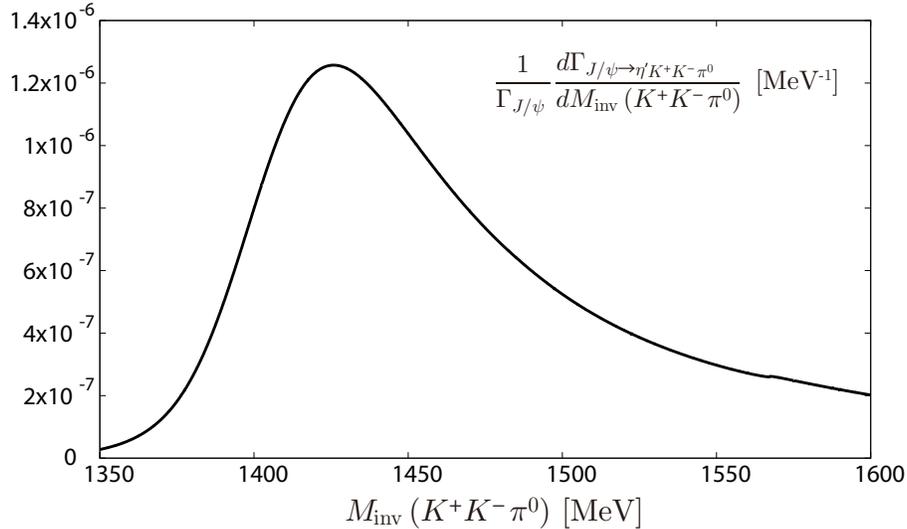}
 \caption{$[d\Gamma/d\minv(K^+K^-\pi^0)]/\Gamma_{J/\psi}$ as a function of $\minv(K^+K^-\pi^0)$ from the integration of Eq.~\eqref{eq:21-2} over $\minv(K^+\pi^0)$.}
 \label{fig:5}
\end{figure}

Next we integrate this differential cross section over $\minv(K^+\pi^0)$
{in the interval $[M_{K^*}-2{\Gamma_{K^*}^0},M_{K^*}+2{\Gamma_{K^*}^0}]$}
and plot it in Fig.~\ref{fig:5}.
What we see in Fig.~\ref{fig:5} is that the peak of $d\Gamma/d\minv(K^+K^-\pi^0)$ is shifted forwards higher energies than the nominal mass of the $h_1(1380)$.
The reason for it is that in Eq.~(\ref{eq:21-2}) there are two factors competing.
One is the $h_1(1380)$ propagator that makes the distribution peak at the nominal mass of the $h_1(1380)$, and the other one is the $K^*$ propagator that becomes more on shell as the energy of $K^{*+}K^-$ increases.
The combination of these factors
makes the peak of the distribution to appear at higher energies.
This explains what is observed in the experiment,
but the mass of the $h_1(1380)$ is the one appearing in its propagator and the measured peak in Ref.~\cite{Besnew} should not be taken as the mass of the $h_1(1380)$.
In other words, our fair reproduction of the mass distribution of Ref.~\cite{Besnew} should be seen as a factor supporting the mass of the $h_1(1380)$
at the nominal PDG mass of $1407~\mev$, which can be observed in other decay channels.

Finally, we integrate $d\Gamma/d\minv(K^+K^-\pi^0)$ over $\minv(K^+K^-\pi^0)$
{in a range of $[M_{K^{*+}}+m_{K^-},M_{J/\psi}-m_{\eta'}]$}
to cover the $h_1(1380)$ resonance seen in Fig.~\ref{fig:5} and multiply by two.

\begin{figure}[t]
 \centering
 \includegraphics[width=12cm]{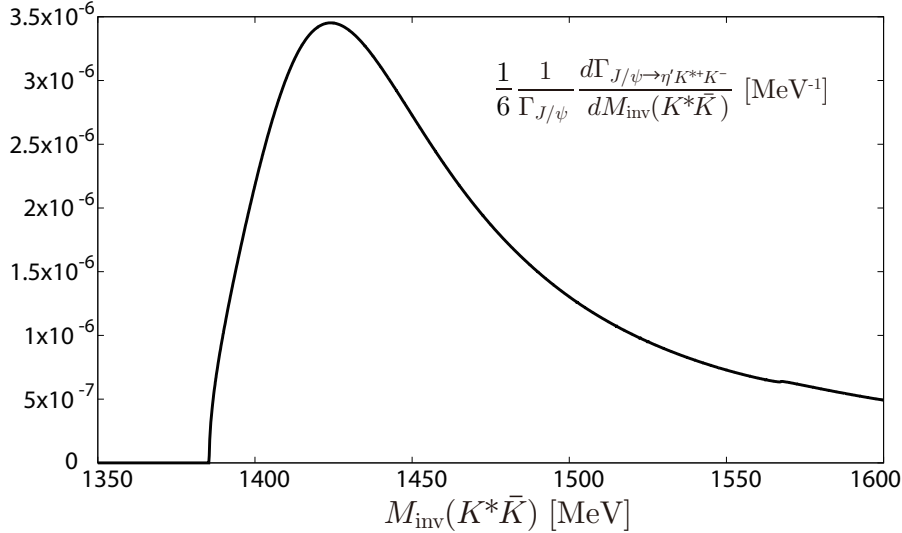}
 \caption{$[d\Gamma/d\minv(K^{*+}K^-)/\Gamma_{J/\psi}]/6$ obtained with Eq.~(\ref{eq:18-1}).}
 \label{fig:6}
\end{figure}

Next
we plot in Fig.~\ref{fig:6} the results
for $d\Gamma/d\minv(K^*\bar{K})$ divided by 6 obtained by means of Eq.~(\ref{eq:18-1})
and we compare it with the results of Fig.~\ref{fig:5} multiplied by two.
Then we integrate $d\Gamma/d\minv(K^*\bar{K})$
and compare these results
with those of the previous method and
the experiment of Ref.~\cite{Besnew}.
Figure~\ref{fig:6} is instructive because there we do not have the $K^*$ propagator, only the $h_1$ one.
In spite of that we observe that the peak of the $K^*\bar{K}$ distribution also appears around $1425~\mev$ as in Fig.~\ref{fig:5}.
Here, the reason for the displacement of the peak versus the nominal mass $1407~\mev$ must be found for the reduced phase space for the $K^{*}\bar{K}$ decay
since the threshold for $K^{*}\bar{K}$ appears at ${1385}~\mev$.
It is then this threshold close to the nominal mass which makes the peak appear at higher energies than the nominal $h_1(1380)$ mass.
Quantitatively we can see that the strength of the mass distribution of Fig.~\ref{fig:5} multiplied by two is a bit smaller than that of Fig.~\ref{fig:6},
and it stretches at lower $K^{*+}K^-$ invariant mass as a consequence of considering explicitly the mass distribution of the $K^*$.
It is rewarding that the two methods give very similar result but we should consider the method that takes into account the $K^*$ mass distribution as more accurate.

When we integrate the distributions over $\minv(K^{*+}K^-)$ {in the range  
{of} $[M_{K^{*+}}+m_{K^-},M_{J/\psi}-m_{\eta'}]$},
we obtain the following
branching ratios 
\begin{description}
 \item[Method 1] $\frac{1}{6}\br(J/\psi\rightarrow\eta'h_1(1380),h_1(1380)\rightarrow K^{*}\bar{K}+c.c.)={4.23\tento{-4}}$
 \item[Method 2] $2\times\br(J/\psi\rightarrow\eta'h_1(1380),h_1(1380)\rightarrow K^{*+}K^-,K^{*+}\rightarrow K^+\pi^0)={3.31\tento{-4}}$
 \item[Experiment] $\br^{\rm exp}(J/\psi\rightarrow\eta'h_1(1380),h_1(1380)\rightarrow K^{*+}K^-+c.c.\ {\rm in}\ K^+K^-\pi^0)\\
	    \hspace{9cm}=(1.51\pm0.09\pm0.21)\tento{-4}$.
\end{description}

As we can see, we obtain with Method 2 a magnitude of $3.31~\tento{-4}$ a bit bigger than the experimental number,
but this should be considered a reasonable success in view that we have not used any free parameters,
and that we admit having uncertainties of the order of $10-20~\%$ in the couplings that revert into uncertainties of $20-49~\%$ in the branching ratios.

As mentioned before, in Table~\ref{tab:ix} we made predictions for other decay models, one of them, the $J/\psi\rightarrow \eta h_1(1380)$,
with a rate substantially smaller than for the other modes and which is tied directly to the assumption made of a dynamically generated $h_1(1380)$ resonance.
Certainly, the measurements of these decay modes will bring relevant information concerning the nature of the axial vector mesons.

\section{Conclusions}
\label{sec:conc}
We have made a study of the $J/\psi\rightarrow\eta'h_1$ and $J/\psi\rightarrow\eta h_1$ decays, with $h_1$ being $h_1(1170)$ and $h_1(1380)$, together with $J/\psi\rightarrow\pi^0b_1(1235)^0$.
For this we have assumed that the axial vector mesons are dynamically generated from the pseudoscalar-vector meson interaction.
Using SU(3) symmetry with a small OZI violating term, which was determined previously in the study of $J/\psi\rightarrow\phi\pi\pi$ and $J/\psi\rightarrow\omega\pi\pi$ reactions,
we are able to obtain rates for these decays, which we compare with experiment.
The comparison with the $J/\psi\rightarrow\pi^0b_1(1235)^0$ data is fair.
On the other hand, the comparison from the recent BESIII experiment \cite{Besnew} requires projection over the $K^*\bar{K}$ channels,
and then selecting $K^{*+}K^-$, $K^{*-}K^+$ and furthermore looking into the $K^+K^-\pi^0$ or $K_S^0K^\pm\pi^\mp$ modes.
We have done that {for the $K^+K^-\pi^0$ mode} using two methods, one where we take the nominal mass of the $K^*$ assuming it as an elementary particle, and another one where the $K^*$ propagator,
accounting for the $K^*$ mass distribution, is explicitly taken into account.
The results obtained with both methods are similar, but the one accounting for the $K^*$ mass distribution is more accurate, and both produce a $K^*\bar{K}$ distribution peaking at higher energies than the nominal $h_1(1380)$ mass, which justifies the results obtained in Ref.~\cite{Besnew}.
Once the integrations over the invariant masses are made, the final results for the branching ratios are in fair agreement with experiment considering errors.
We also make predictions for modes not yet observed.
We should stress that, apart from some uncertainty in the input used, from the study of the $J/\psi\rightarrow\phi\pi\pi$, $\omega\pi\pi$ reactions,
where no axial vector mesons were produced, we have no freedom in our approach.
The predictions we made are tied to the nature assumed for the axial vector meson as dynamically generated from the pseudoscalar-vector meson interaction,
and agreement with experiment should be seen as a factor in favor of this hypothesis.
The investigation of the modes studied here, not yet measured, should be encouraged in this context.

\begin{acknowledgments}
This work is partly supported by the National Natural Science Foundation of China (Grants No. 11565007 and 11847317).
This work is also partly supported by the Spanish Ministerio de Economia y Competitividad
and European FEDER funds under the contract number FIS2011-28853-C02-01, FIS2011-28853-C02-02, FIS2014-57026-REDT, FIS2014-51948-C2-1-P, and FIS2014-51948-C2-2-P, and the Generalitat Valenciana in the program Prometeo II-2014/068.
S.~Sakai acknowledges the support by NSFC and DFG through funds provided to the Sino-German CRC110 ``Symmetries and the Emergence of Structure in QCD'' (NSFC Grant No.~11621131001), by the NSFC (Grant No.~11747601), by the CAS Key Research Pro-gram of Frontier Sciences (Grant No.~QYZDB-SSW-SYS013) and by the CAS Key Research Program (Grant No.~XDPB09).
\end{acknowledgments}



\begin{thebibliography}{99}

\bibitem{isgur}
  S.~Godfrey and N.~Isgur,
  Phys.\ Rev.\ D {\bf 32}, 189 (1985).

\bibitem{vijande}
  J.~Vijande, F.~Fernandez and A.~Valcarce,
  J.\ Phys.\ G {\bf 31}, 481 (2005).

\bibitem{Pelaez}
  J.~R.~Pelaez,
  Phys.\ Rept.\  {\bf 658} (2016) 1.

\bibitem{rocaram}
  L.~S.~Geng, E.~Oset, J.~R.~Pelaez and L.~Roca,
  Eur.\ Phys.\ J.\ A {\bf 39}, 81 (2009).

\bibitem{Lutz}
  M.~F.~M.~Lutz and E.~E.~Kolomeitsev,
  Nucl.\ Phys.\ A {\bf 730}, 392 (2004).

\bibitem{Luis}
  L.~Roca, E.~Oset and J.~Singh,
  Phys.\ Rev.\ D {\bf 72}, 014002 (2005).

\bibitem{geng}
  Y.~Zhou, X.~L.~Ren, H.~X.~Chen and L.~S.~Geng,
  Phys.\ Rev.\ D {\bf 90}, 014020 (2014).

\bibitem{Leupold}
  M.~Wagner and S.~Leupold,
  Phys.\ Rev.\ D {\bf 78}, 053001 (2008).

\bibitem{Birse}
  M.~C.~Birse,
  Z.\ Phys.\ A {\bf 355}, 231 (1996).

\bibitem{rocahosa}
  L.~Roca, A.~Hosaka and E.~Oset,
  Phys.\ Lett.\ B {\bf 658}, 17 (2007).

\bibitem{hiderroca}
  H.~Nagahiro, L.~Roca and E.~Oset,
  Phys.\ Rev.\ D {\bf 77}, 034017 (2008).

\bibitem{yongseok}
  K.~S.~Jeong, S.~H.~Lee and Y.~Oh,
  JHEP {\bf 1808}, 179 (2018).

\bibitem{Lutzleo}
  M.~F.~M.~Lutz and S.~Leupold,
  Nucl.\ Phys.\ A {\bf 813}, 96 (2008).

\bibitem{hiddenaga}
  H.~Nagahiro, L.~Roca, A.~Hosaka and E.~Oset,
  Phys.\ Rev.\ D {\bf 79}, 014015 (2009).

\bibitem{Roig}
  I.~M.~Nugent, T.~Przedzinski, P.~Roig, O.~Shekhovtsova and Z.~Was,
  Phys.\ Rev.\ D {\bf 88}, 093012 (2013).

\bibitem{Volkovtau}
  M.~K.~Volkov and K.~Nurlan,
  Phys.\ Part.\ Nucl.\ Lett.\  {\bf 14}, 677 (2017).


\bibitem{Osipov}
  A.~A.~Osipov,
  arXiv:1812.06476 [hep-ph].

\bibitem{Volkovtauf1}
  A.~V.~Vishneva, M.~K.~Volkov and D.~G.~Kostunin,
  Eur.\ Phys.\ J.\ A {\bf 50}, 137 (2014).

\bibitem{luistauf1}
  E.~Oset and L.~Roca,
  Phys.\ Lett.\ B {\bf 782}, 332 (2018).

\bibitem{Dairoca}
  L.~R.~Dai, L.~Roca and E.~Oset,
  arXiv:1811.06875 [hep-ph].

\bibitem{calderon}
  H.~Y.~Cheng, C.~K.~Chua, K.~C.~Yang and Z.~Q.~Zhang,
  Phys.\ Rev.\ D {\bf 87}, 114001 (2013).

\bibitem{Molina}
  R.~Molina, M.~D\"{o}ring and E.~Oset,
  Phys.\ Rev.\ D {\bf 93}, 114004 (2016).

\bibitem{ZhangXie}
  X.~Zhang and J.~J.~Xie,
  arXiv:1812.04242 [hep-ph].

\bibitem{OsiVolkov}
  A.~A.~Osipov, A.~A.~Pivovarov and M.~K.~Volkov,
  Phys.\ Rev.\ D {\bf 98}, 014037 (2018).

\bibitem{Aceti}
  F.~Aceti, J.~J.~Xie and E.~Oset,
  Phys.\ Lett.\ B {\bf 750}, 609 (2015).

\bibitem{PDG}
M. Tanabashi {\it et al.} (Particle Data Group), Phys.\ Rev.\ D {\bf 98}, 030001 (2018).

\bibitem{Acetiso}
  F.~Aceti, J.~M.~Dias and E.~Oset,
  Eur.\ Phys.\ J.\ A {\bf 51}, 48 (2015).

\bibitem{Oller}
  J.~A.~Oller and E.~Oset,
  Nucl.\ Phys.\ A {\bf 620}, 438 (1997);
  Erratum: [Nucl.\ Phys.\ A {\bf 652}, 407 (1999)].

\bibitem{Kaiser}
  N.~Kaiser,
  Eur.\ Phys.\ J.\ A {\bf 3}, 307 (1998).

\bibitem{Markushin}
  M.~P.~Locher, V.~E.~Markushin and H.~Q.~Zheng,
  Eur.\ Phys.\ J.\ C {\bf 4}, 317 (1998).

\bibitem{juan}
  J.~Nieves and E.~Ruiz Arriola,
  Nucl.\ Phys.\ A {\bf 679}, 57 (2000).

\bibitem{BESiso}
  M.~Ablikim {\it et al.} [BESIII Collaboration],
  Phys.\ Rev.\ D {\bf 92}, 012007 (2015).

\bibitem{cao}
  C.~Cheng, J.~J.~Xie and X.~Cao,
  Commun.\ Theor.\ Phys.\  {\bf 66},  675 (2016).

\bibitem{Volkovee}
  M.~K.~Volkov, A.~A.~Pivovarov and A.~A.~Osipov,
  Int.\ J.\ Mod.\ Phys.\ A {\bf 32},  1750123 (2017).

\bibitem{XieLam}
  J.~J.~Xie,
  Phys.\ Rev.\ C {\bf 92}, 065203 (2015).

\bibitem{CLASf1}
  R.~Dickson {\it et al.} [CLAS Collaboration],
  Phys.\ Rev.\ C {\bf 93}, 065202 (2016).

\bibitem{He}
  X.~Y.~Wang and J.~He,
  Phys.\ Rev.\ D {\bf 95},  094005 (2017).

\bibitem{Jujun}
  J.~J.~Xie and E.~Oset,
  Phys.\ Lett.\ B {\bf 753}, 591 (2016).

\bibitem{besalba}
  M.~Ablikim {\it et al.} [BES Collaboration],
  Phys.\ Lett.\ B {\bf 685}, 27 (2010).


\bibitem{albala}
  J.~J.~Xie, M.~Albaladejo and E.~Oset,
  Phys.\ Lett.\ B {\bf 728}, 319 (2014).

\bibitem{GengOset}
  L.~S.~Geng and E.~Oset,
  Phys.\ Rev.\ D {\bf 79}, 074009 (2009).

\bibitem{Cabrera}
  D.~Cabrera, D.~Jido, R.~Rapp and L.~Roca,
  Prog.\ Theor.\ Phys.\  {\bf 123}, 719 (2010).

\bibitem{Besnew}
  M.~Ablikim {\it et al.} [BESIII Collaboration],
  Phys.\ Rev.\ D {\bf 98},  072005 (2018).

\bibitem{ulfoller}
  U.~G.~Mei{\ss}ner and J.~A.~Oller,
  Nucl.\ Phys.\ A {\bf 679}, 671 (2001).

\bibitem{Palomar}
  L.~Roca, J.~E.~Palomar, E.~Oset and H.~C.~Chiang,
  Nucl.\ Phys.\ A {\bf 744}, 127 (2004).

\bibitem{chic1}
  W.~H.~Liang, J.~J.~Xie and E.~Oset,
  Eur.\ Phys.\ J.\ C {\bf 76}, 700 (2016).

\bibitem{midhalo}
  M.~Ablikim {\it et al.} [BESIII Collaboration],
  Phys.\ Rev.\ D {\bf 95},  032002 (2017).

\bibitem{etac}
  V.~R.~Debastiani, W.~H.~Liang, J.~J.~Xie and E.~Oset,
  Phys.\ Lett.\ B {\bf 766}, 59 (2017).

\bibitem{Bramon}
  A.~Bramon, A.~Grau and G.~Pancheri,
  Phys.\ Lett.\ B {\bf 283}, 416 (1992).

\bibitem{rocageng}
  L.~S.~Geng, E.~Oset, L.~Roca and J.~A.~Oller,
  Phys.\ Rev.\ D {\bf 75}, 014017 (2007).

\bibitem{Sakai}
  R.~Pavao, S.~Sakai and E.~Oset,
  Eur.\ Phys.\ J.\ C {\bf 77},  599 (2017).

  \end{thebibliography}
  \end{document}